\documentclass[american,prb,twocolumn,twoside,superscriptaddress,showpacs]{revtex4}
\usepackage[T1]{fontenc}
\usepackage[latin9]{inputenc}
\usepackage{amsthm}
\usepackage{amsmath}
\usepackage{graphicx}
\usepackage{amssymb}
\usepackage{esint}
\usepackage{babel}

\begin{document}

\title{Molecular dynamics study of the thermopower of Ag, Au, and Pt nanocontacts}

\author{F. Pauly}

\affiliation{Institut f\"ur Theoretische Festk\"orperphysik and DFG Center for
Functional Nanostructures, Karlsruhe Institute of Technology, 76131
Karlsruhe, Germany}

\email{fabian.pauly@kit.edu}

\author{J. K. Viljas}

\affiliation{Low Temperature Laboratory, Aalto University, P.O. Box 15100, 00076
Aalto, Finland}

\author{M. B\"urkle}

\affiliation{Institut f\"ur Theoretische Festk\"orperphysik and DFG Center for
Functional Nanostructures, Karlsruhe Institute of Technology, 76131
Karlsruhe, Germany}

\author{M. Dreher}

\affiliation{Fachbereich Physik, Universit\"at Konstanz, 78457 Konstanz, Germany}

\author{P. Nielaba}

\affiliation{Fachbereich Physik, Universit\"at Konstanz, 78457 Konstanz, Germany}

\author{J. C. Cuevas}

\affiliation{Departamento de F\'isica Te\'orica de la Materia Condensada, Universidad
Aut\'onoma de Madrid, 28049 Madrid, Spain}
\begin{abstract}
Using molecular dynamics simulations of many junction stretching processes
we analyze the thermopower of silver (Ag), gold (Au), and platinum
(Pt) atomic contacts. In all cases we observe that the thermopower
vanishes on average within the standard deviation and that its fluctuations
increase for decreasing minimum cross-section of the junctions. However,
we find a suppression of the fluctuations of the thermopower for the
$s$-valent metals Ag and Au, when the conductance originates from
a single, perfectly transmitting channel. Essential features of the
experimental results for Au, Ag, and copper (Cu) of Ludoph\emph{ }\textit{\emph{and
van Ruitenbeek}} {[}Phys.\ Rev.\ B \textbf{59}, 12290 (1999){]},
as yet unaddressed by atomistic studies, can hence be explained by
considering the atomic and electronic structure at the disordered
narrowest constriction of the contacts. For the multivalent metal
Pt our calculations predict the fluctuations of the thermopower to
be larger by one order of magnitude as compared to Ag and Au, and
suppressions of the fluctuations as a function of the conductance
are absent.
\end{abstract}


\maketitle

\section{Introduction}

The development of thermoelectric devices for the efficient conversion
of heat into electrical energy or for refrigeration would be an important
step towards a more environmentally friendly use of energy. Engineered
nanostructures are promising materials in this respect,\cite{Venkatasubramanian:Nature2001,Boukai:Nature2008,Hochbaum:Nature2008,Hochbaum:ChemRev2010}
and molecular junctions are presently moving into the focus of research.\cite{Malen:CPL2010,Nozaki:PRB2010,Dubi:RevModPhys2011}
They seem to be good candidates for achieving high thermoelectric
figures of merit as a result of the discrete energy level structure
of the molecules\cite{Mahan:PNAS1996,Malen:CPL2010} and because molecular
properties can be controlled by chemical synthesis. Already the measurement
of the thermopower alone provides important insights into the electronic
structure of molecular junctions, not accessible by conductance measurements.\cite{Reddy:Science2007}
Thus, electron and hole conduction can be distinguished\cite{Ashcroft:1976}
and the alignment of molecular levels with respect to the Fermi energy
can be determined.\cite{Reddy:Science2007,Baheti:NL2008,Malen:NL2009Length}
It turns out that the experiments are described by a combination of
electronic structure and transport calculations.\cite{Pauly:PRB2008,Ke:NL2009}
For instance, the linear increase of the thermopower with molecule
length in the typical off-resonant transport situation has been found
in both experiment\cite{Reddy:Science2007,Malen:NL2009Length} and
theory.\cite{Segal:PRB2005,Viljas:PRB2008everything,Pauly:PRB2008}

While molecular junctions appear promising from a device-oriented
point of view, metallic electrodes are typically used to contact the
molecules.\cite{Reddy:Science2007,Baheti:NL2008,Malen:NL2009Length}
Metallic atomic contacts hence constitute important reference systems.
Until now their thermopower has been studied experimentally in the
single-atom contact regime only by Ludoph and van Ruitenbeek.\cite{Ludoph:PRB1999}
The authors discuss mostly their results for Au contacts, but measurements
for Ag and Cu contacts have been performed as well.\cite{Ludoph:PRB1999,Ludoph:PhD1999}

Motivated by pioneering experiments with high-mobility two-dimensional
electron gases,\cite{Molenkamp:PRL1990,Houten:SST1992} theoretical
studies have described the thermopower of quantum point contacts in
two and three dimensions using the free-electron-gas approximation
for adiabatic constrictions.\cite{Streda:JPhysCondMat1989,Houten:SST1992,Bogacheck:PRB1996,Kokurin:JPCM2004}
Indeed, such studies have successfully predicted maxima of the thermopower
at the transition between quantized conductance plateaus.\cite{Streda:JPhysCondMat1989,Molenkamp:PRL1990}
The Fermi wavelength of several 10 nm for the two-dimensional electron
gases (see Ref.~\citenum{vanWees:PRL1998} for instance) is much
larger than the atomic dimensions. For metals, in contrast, the Fermi
wavelength is on the order of the interatomic distance.\cite{Ashcroft:1976}
Therefore, disorder-related effects due to the atomic structure are
expected to play a more important role than for the two-dimensional
electron gas devices, and material-specific chemical properties enter
as an additional aspect. To describe the metallic atomic contacts,
it is hence necessary to take into account both the atomic and the
detailed electronic structure.\cite{Cuevas:PRL1998a,Scheer:Nature1998,Ludoph:PRB2000,Brandbyge:PRB2002,Agrait:PhysRep2003,Dreher:PRB2005,Pauly:PRB2006,Pauly:NJP2008}

Recent theoretical studies of the thermopower of nanocontacts have
considered separately the effects of disorder and material-specific
parameters. Thus, thermoelectric effects for structures with atomically-thin
one-dimensional wires connected to two-dimensional electrodes were
studied in Refs.~\citenum{Dubi:NanoLett2009,Dubi:PRB2009} for various
geometries and with disorder, but without material-specific parameters.
In contrast in Refs.~\citenum{Wang:PRB2005,Liu:ACSN2009} material-specific
parameters were employed, while crucial disorder-related effects,
resulting in the statistical nature of the experiments,\cite{Ludoph:PRB1999}
were neglected by treating only ideal wire geometries.

Based on molecular dynamics (MD) simulations of many junction stretching
processes, we study here the thermopower of atomic contacts for three
different metals, namely Ag, Au, and Pt. We use an atomistic model,
a material-specific description of the electronic structure, and address
the influence of disorder at the narrowest part of the contacts. The
results for the monovalent metals, especially Au, are compared to
the available experimental data,\cite{Ludoph:PRB1999} while predictions
are made for the multivalent Pt.

The manuscript is organized as follows. In Sec.~\ref{sec:Approach}
we present the methodological details of our atomistic simulations
and then discuss the results in Sec.~\ref{sec:Results}. Conclusions
and an outlook are given in Sec.~\ref{sec:Conclusions}. Additional
material, such as further examples of junction stretching processes
with animations or technical details, are provided in the appendices
and in the electronic supplemental material.\cite{EPAPS}

\section{Theoretical approach\label{sec:Approach}}

Our calculations proceed along the lines of Refs.~\citenum{Dreher:PRB2005,Pauly:PRB2006},
which we extend here to study the thermopower. The calculations can
be divided essentially into two parts, namely the generation of contact
geometries and the determination of transport properties.

\subsection{Molecular dynamics simulations}

The contact geometries used in this work for Ag, Au, and Pt are those
employed in our previous studies.\cite{Dreher:PRB2005,Pauly:PRB2006}
For the sake of completeness, we describe their construction briefly.

We determine contact geometries by performing classical MD simulations
of junction stretching processes. As displayed in Figs.~\ref{fig:Ag2-contact}
to \ref{fig:Pt2-contact},
we choose nanowire geometries with a central wire (CW) connected to
larger-diameter electrodes which are attached at the top and bottom.
Initially, we assume all atoms to be located at the positions of a
perfect fcc lattice, with the $\left\langle 100\right\rangle $ direction
oriented along the $z$ axis. The lattice constants are determined
by minimizing the potential energy of a crystal. When we use the semiempirical
potentials of our MD calculations,\cite{Jacobsen:SurfSci1996} we
obtain lattice constants of 0.408, 0.406, 0.393 nm for Ag, Au, Pt,
respectively, which are close to the experimental values.\cite{Ashcroft:1976}
The CW consists of 112 atoms with 14 layers along the $z$ direction
and 8 atoms per layer. Its initial length amounts to 2.65, 2.64, 2.55
nm for Ag, Au, Pt, respectively, as indicated in Figs.~\ref{fig:Ag2-contact}
to \ref{fig:Pt2-contact}.
In each case the electrodes at the top and bottom contain 288 atoms. 

Junction stretching is performed by separating the otherwise fixed
electrodes symmetrically by a constant distance during every time
step of 1.4 fs. In this process, we use periodic boundary conditions
along the $z$ direction and the minimum image convention for the
supercells perpendicular to it.\cite{Allen:1989} The constant stretching
speed amounts to 2 m/s. The forces on the wire atoms are calculated
from semiempirical potentials,\cite{Jacobsen:SurfSci1996} while their
average temperature remains at 4.2 K by use of a Nos\'e-Hover thermostat.\cite{Frenkel:2001}
Different junction evolutions are obtained by choosing random starting
velocities for the atoms in the CW. Every 1.4 ps a contact geometry
is recorded, and the strain force is calculated.\cite{Finbow:MolPhys1997}

\subsection{Transport properties}

Transport properties are determined within the Landauer-B\"uttiker
formalism.\cite{Houten:SST1992,Dreher:PRB2005,Pauly:PRB2006,Pauly:PRB2008,Pauly:NJP2008,Cuevas:2010}
The conductance $G$ and thermopower $S$ are expressed as\begin{equation}
G=G_{0}K_{0}(T),\label{eq:G}\end{equation}
\begin{equation}
S=-\frac{K_{1}(T)}{eTK_{0}(T)},\label{eq:S}\end{equation}
with $K_{n}(T)=\int dE\left(E-\mu\right)^{n}\tau(E)[-\partial_{E}f(E,T)]$,
the quantum of conductance $G_{0}=2e^{2}/h$, the absolute value of
the electron charge $e=\left|e\right|$, the transmission function
$\tau(E)$, the Fermi function $f(E,T)=\{\exp[(E-\mu)/k_{B}T]+1\}^{-1}$,
the Boltzmann constant $k_{B}$, and the electrochemical potential
$\mu$, which approximately equals the metal Fermi energy $\mu\approx E_{F}$.

In order to compare to the experiments of Ref.~\citenum{Ludoph:PRB1999},
we assume in the following a temperature of $T=12$ K for the determination
of the transport properties. While we evaluate $S$ via Eq.~(\ref{eq:S}),
for the conductance the simpler low-temperature formula \begin{equation}
G=G_{0}\tau(E_{F})=G_{0}\sum_{n}\tau_{n}(E_{F})\label{eq:GlowT}\end{equation}
yields a good approximation. Here, $\tau$ has been resolved into
the contributions of transmission probabilities $\tau_{n}$ from the
individual transmission eigenchannels $n$. As discussed below, they
provide important information for the understanding of the results.

We obtain the transmission and the channel decomposition by use of
Green's function techniques, as described in Ref.~\citenum{Pauly:PRB2006}.
We assume the central device region to be equal to the CW, while the
remaining atoms of the contacts are attributed to the electrodes.
However, while in Ref.~\citenum{Pauly:PRB2006} we assumed the unperturbed
electrode Green's functions to be those of the bulk, in the present
study we use surface Green's functions. They are determined via a
decimation technique by modeling the electrodes as surfaces of ideal,
semi-infinite crystals.\cite{Guinea:PRB1983,Pauly:NJP2008} Even though
the results for these two different procedures differ only slightly,\cite{Pauly:PhD2007}
surface Green's functions describe the physical situation more accurately.

To perform the energy integrations in Eq.~(\ref{eq:S}), we have computed
$\tau(E)$ every 5.6 ps for 11 equally spaced points in the energy
interval $\left[E_{F}-\Delta,E_{F}+\Delta\right]$ with $\Delta=0.05$
eV. $K_{n}(T)$ is then obtained by the integration of $\left(E-\mu\right)^{n}\tau(E)\left[-\partial_{E}f(E,T)\right]$,
using a linear interpolation for $\tau(E)$ between the energy sampling
points.

The effective single-particle Hamiltonian and overlap matrices for
the evaluation of the transmission are obtained from a Slater-Koster\cite{Slater:PR1954}
tight-binding description,\cite{Mehl:PRB1996,Mehl:Book1998} supplemented
by a local charge neutrality condition.\cite{Pauly:PRB2008} Even
if this approach is not at the level of self-consistent ab-initio
methods, it is still atomistic and takes into account the symmetries
of the $s$, $p$, and $d$ valence orbitals for these monoatomic
systems. In the past, it was used successfully to describe the conduction
properties of various metallic atomic contacts.\cite{Dreher:PRB2005,Viljas:PRB2005,Pauly:PRB2006,Viljas:PRB2007,Hafner:PRB2008}

\section{Results and discussion\label{sec:Results}}

\subsection{Junction stretching events}

\begin{figure}[!tb]
\begin{centering}
\includegraphics[width=1\columnwidth]{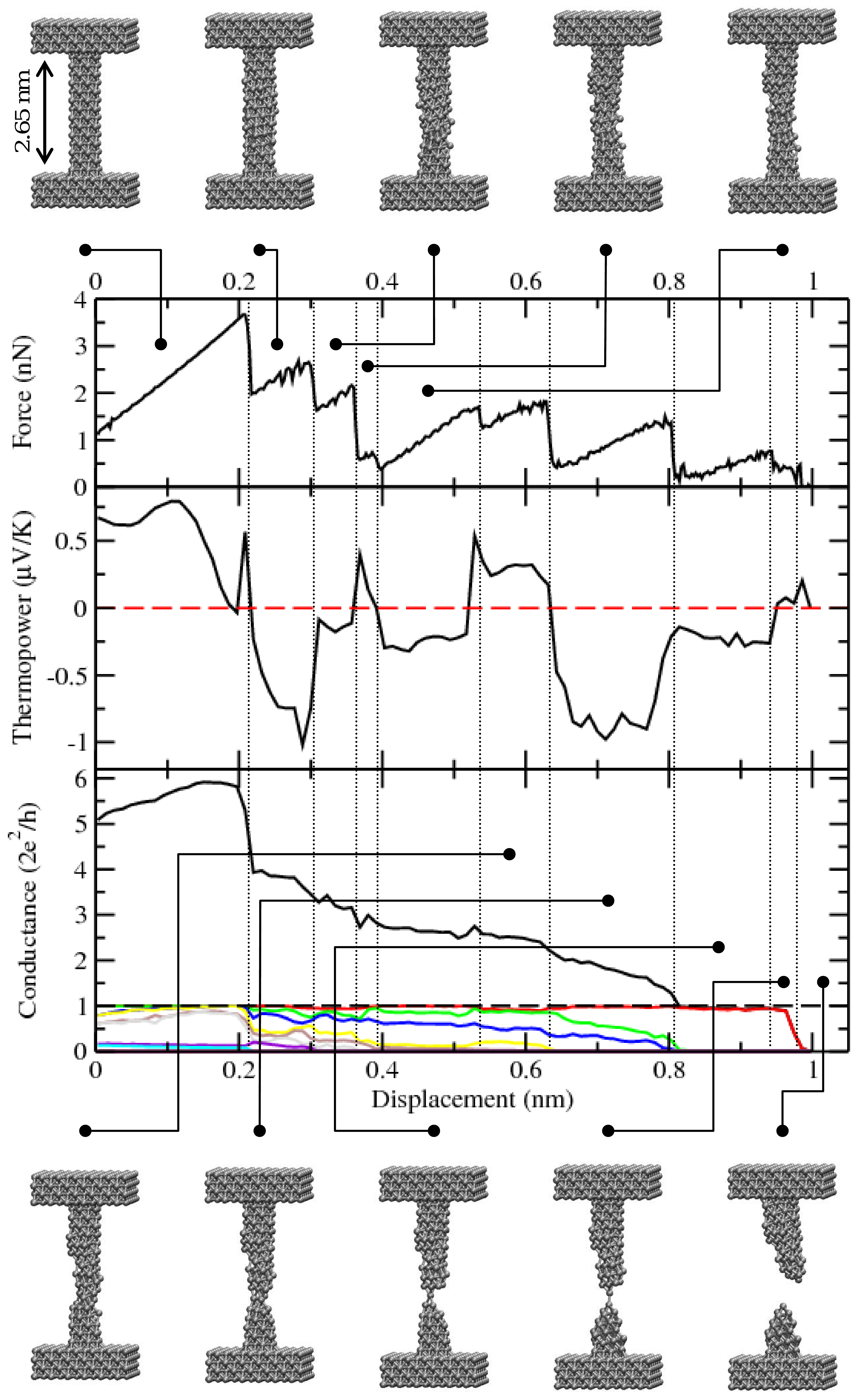}
\par\end{centering}

\caption{\label{fig:Ag2-contact}(Color online) Formation of an Ag dimer contact.
The upper, middle, and lower panel shows, respectively, the tensile
force, the thermopower, and the conductance as a function of the electrode
displacement. For the conductance also the decomposition into conduction
eigenchannels $G_{0}\tau_{n}$ is displayed. Horizontal dashed lines
indicate the zero level of the thermopower or a conductance of $1G_{0}$,
while the vertical dotted lines mark the main plastic reorganizations
of the contact. Above and below the graphs snapshots of the contact
geometry during the stretching process are shown, and the length of
the central wire is given for zero electrode displacement.}
\end{figure}

\begin{figure}[!tb]
\begin{centering}
\includegraphics[width=1\columnwidth]{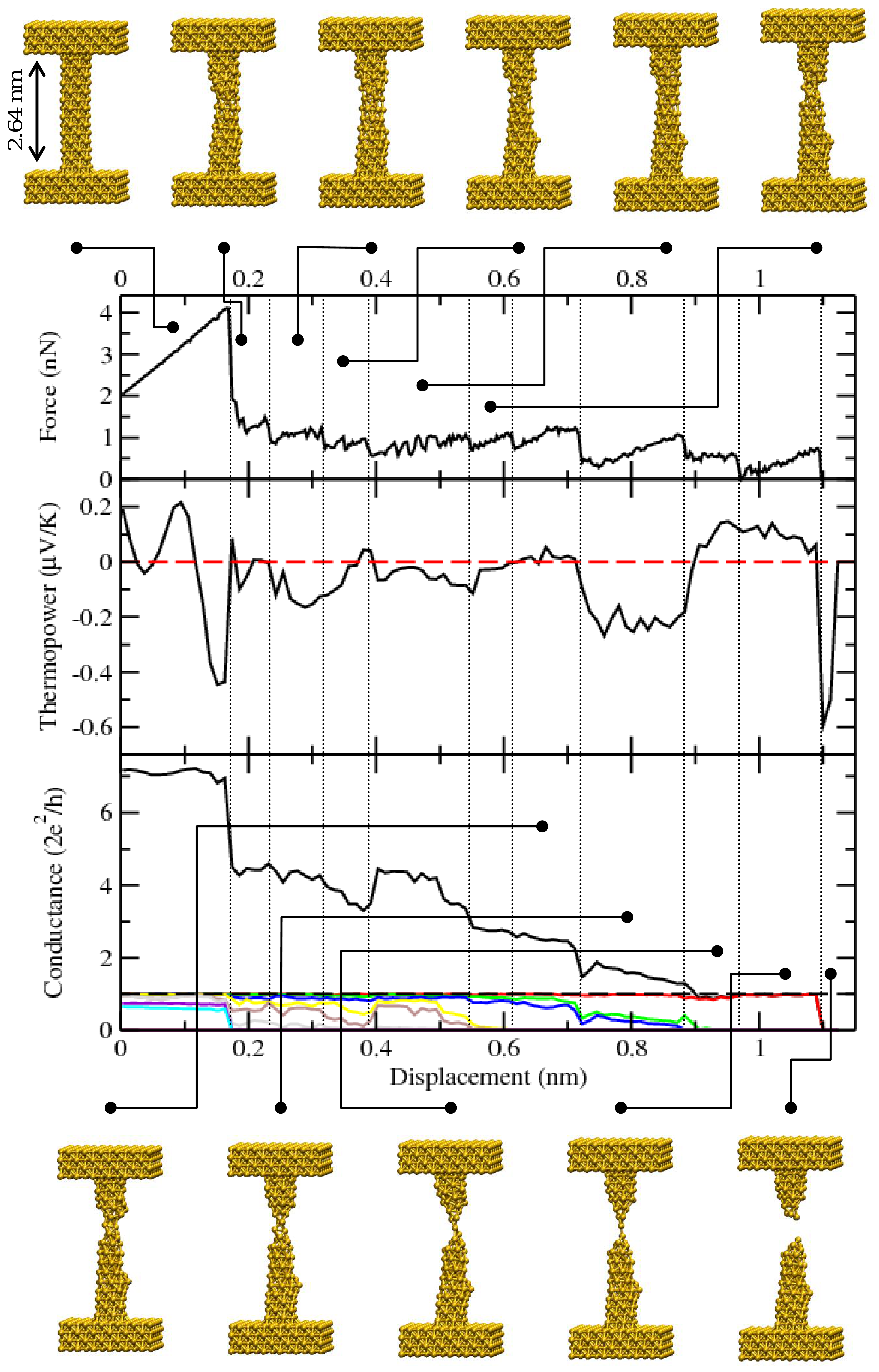}
\par\end{centering}

\caption{\label{fig:Au2-contact}(Color online) The same as Fig.~\ref{fig:Ag2-contact},
but for the formation of an Au dimer contact.}
\end{figure}

\begin{figure}[!tb]
\begin{centering}
\includegraphics[width=1\columnwidth]{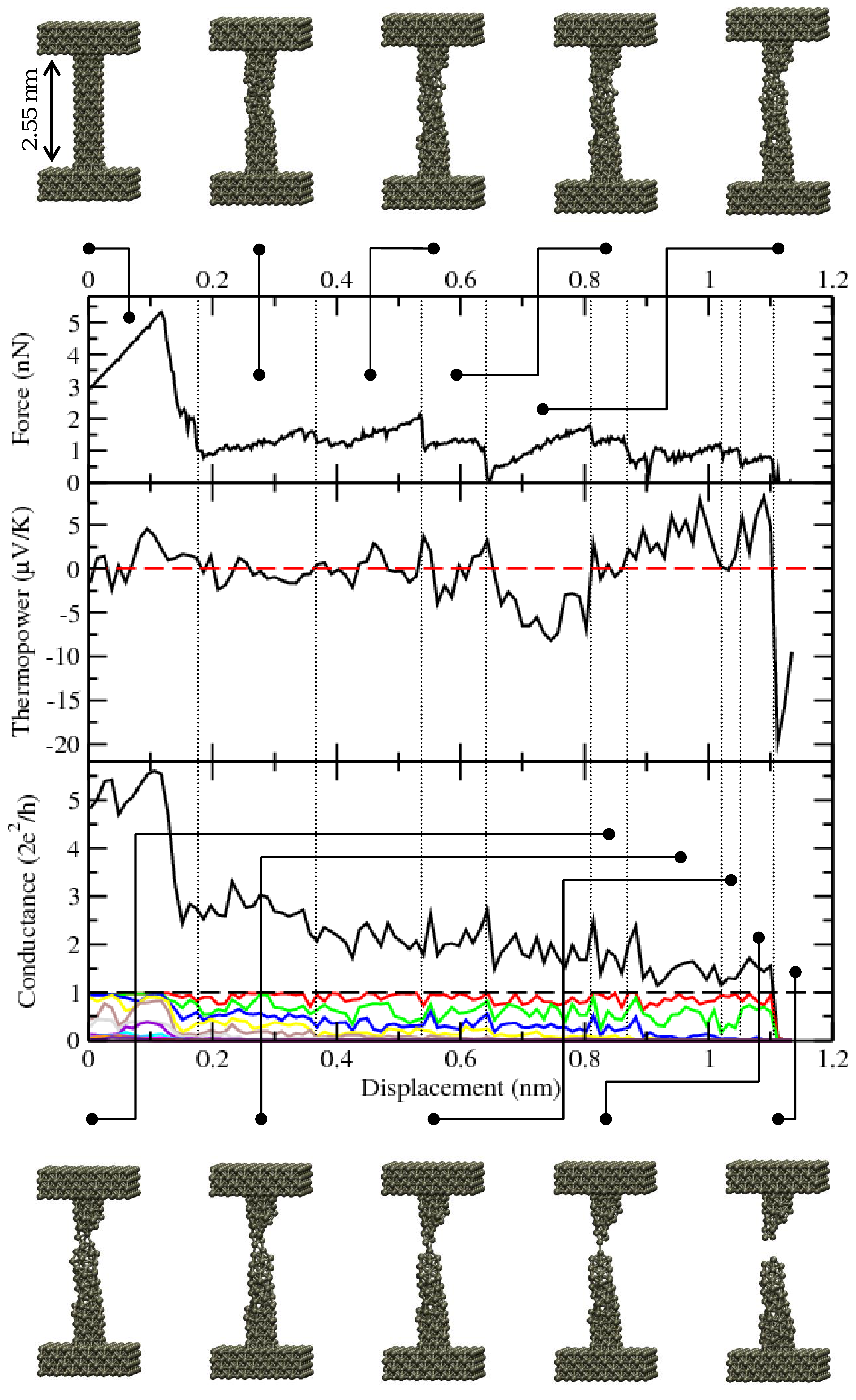}
\par\end{centering}

\caption{\label{fig:Pt2-contact}(Color online) The same as Fig.~\ref{fig:Ag2-contact},
but for the formation of a Pt dimer contact.}
\end{figure}

In order to analyze the behavior of the thermopower for the metallic
atomic contacts, we have simulated 50 stretching processes for each
of the Ag, Au, and Pt nanowires. Beside the thermopower we have analyzed
the strain force and the conductance with its decomposition into individual
channel contributions. Examples of stretching events, which we will
discuss in the following paragraphs, are shown in Figs.~\ref{fig:Ag2-contact}
to \ref{fig:Pt2-contact}.
In all the selected cases the contacts break after a dimer contact
has formed, i.e.\ a two-atom chain. Further examples of stretching
events, including the formation of longer atomic chains for Au and
Pt contacts, are provided in Appendix \ref{sec:stretchAdd}.

The evolution of the mechanical and transport properties for a Ag
contact are displayed in Fig.~\ref{fig:Ag2-contact}.%
 From the strain force, shown in the upper panel of the figure, elastic
and plastic stages can be distinguished. While the force increases
in a linear manner in the elastic ones, it drops suddenly when bonds
break and atoms rearrange during the short plastic stages.\cite{Robio-Bollinger:PRL2001}
Similar to what is shown in the experimental plots for Au in Ref.~\citenum{Ludoph:PRB1999},
the thermopower in the middle panel behaves in a step-wise manner.
It takes both positive and negative values, and fluctuates around
zero. As visible from the dotted lines in Fig.~\ref{fig:Ag2-contact},
the steps in the thermopower typically coincide with plastic deformations
of the contact with smooth changes in between. All these features
and also the absolute values of the thermopower are in agreement with
the experimental results. We note that while data are presented in
Ref.~\citenum{Ludoph:PRB1999} only for Au, it is stated there that
studies of Ag and Cu samples showed similar results, which justifies
our comparison. The conductance is displayed in the lowest panel of
Fig.~\ref{fig:Ag2-contact}. After an initial increase, it drops in a
gradual manner. Before contact rupture it decreases from a value of
around 2$G_{0}$ to a clear plateau at $1G_{0}$, when a single atom
is at the narrowest constriction. The contact breaks quickly after
a dimer has formed. For the single-atom and dimer configurations the
current is carried by a single fully transmitting channel, as expected
for the $s$-valent metals.\cite{Scheer:Nature1998,Ludoph:PRL1999,Ludoph:PRB2000,Dreher:PRB2005,Pauly:PRB2006}

The evolution of junction properties for a Au contact is shown in
Fig.~\ref{fig:Au2-contact}.%
 As suggested by the comparable valence electronic structure of the
atoms, the results resemble those of Ag. The thermopower behaves in
a step-wise fashion and exhibits gradual changes or plateaus during
the elastic stages, interrupted by the plastic deformations. Again
it takes both positive and negative values, and the thermopower is
comparable in magnitude to those of Ag and the data presented in Ref.~\citenum{Ludoph:PRB1999}.
The conductance falls with a revival to a value of $1.9G_{0}$. It
decreases then in a rather continuous way to a value of $1.3G_{0}$
before it drops to a value slightly below $1G_{0}$, carried by a
single channel. The conductance changes only little, when the single-atom
contact transforms into the rather stable dimer contact.

For multivalent Pt the situation is expected to change, and an example
is shown in Fig.~\ref{fig:Pt2-contact}.%
 We observe that also here the thermopower fluctuates around zero,
but with absolute values roughly one order of magnitude larger as
compared to Ag and Au. The occurrence of these strong fluctuations,
seen also in the conductance, is due to the important contribution
of $d$ states at $E_{F}$. They give rise to a pronounced energy
dependence of the transmission especially for the narrow contacts,
as discussed in Ref.~\citenum{Pauly:PRB2006}. Also, due to the higher
anisotropy of these states as compared to the spherically symmetric
$s$ valence orbitals of Ag and Au the junction transport properties
exhibit a high sensitivity to changes in the atomic positions.\cite{Pauly:PRB2006}
While we are not aware of any measurements of the thermopower for
atomic Pt contacts, we observe that the conductance for the single-atom
and dimer contacts is at around $1.5G_{0}$ (see the electrode displacements
above $1.02$ nm). The multivalent electronic structure and hence
the contribution of several channels to the transmission at $E_{F}$
typically lead to conductance values above $1G_{0}$ for one-atom-wide
constrictions, in agreement with the interpretation of measured opening
traces and conductance histograms.\cite{Scheer:Nature1998,Yanson:PhD2001,Smit:Nature2002,Kiguchi:PRL2008}

\subsection{Statistical analysis of atomistic simulations}

Collecting the data of many stretching processes, we perform a statistical
analysis similar to the experiments.\cite{Ludoph:PRB1999} For this
purpose, we use the 50 simulated opening events of the nanowires for
each metal. Typically the contacts break via the single atom (Ag:
12\%; Au: 8\%; Pt: 4\%) or dimer (Ag: 82\%; Au: 68\%; Pt: 60\%) configurations
mentioned before, with a single atom or a chain of two atoms in the
narrowest part, respectively. As visible from the percentage of occurrence,
the dimers are generally preferred. However, we find also atomic chain
geometries with chain lengths of three atoms or more (Ag: 6\%; Au:
24\%; Pt: 36\%). For Ag such structures are rare. Out of the 50 stretching
events, we find at rupture only three chain configurations with chain
lengths varying from three to five atoms. While such structures are
generally believed to be unlikely,\cite{Smit:PRL2001,Bahn:PRL2001}
they have been observed in transmission electron microscopy studies.\cite{Rodrigues:PRB2002}
For Au and Pt atomic chains occur more frequently in our simulations,
and they are longer. We have decided to exclude the junctions breaking
with chains of 10 atoms or more in length from our analysis, since
experiments indicate maximum lengths of up to eight atoms.\cite{Yanson:Nature1998,Yanson:PhD2001,Agrait:PRL2002}
For Au we have additionally excluded an opening process leading to
a chain with six atoms, because the thermopower exhibits peculiar
features before contact rupture related to $d$ states. For this reason
we present below the statistical analysis of the transport properties
with 50 stretching events for Ag, 46 for Au, and 41 for Pt. 

\begin{figure*}[!t]
\begin{centering}
\includegraphics[width=2\columnwidth]{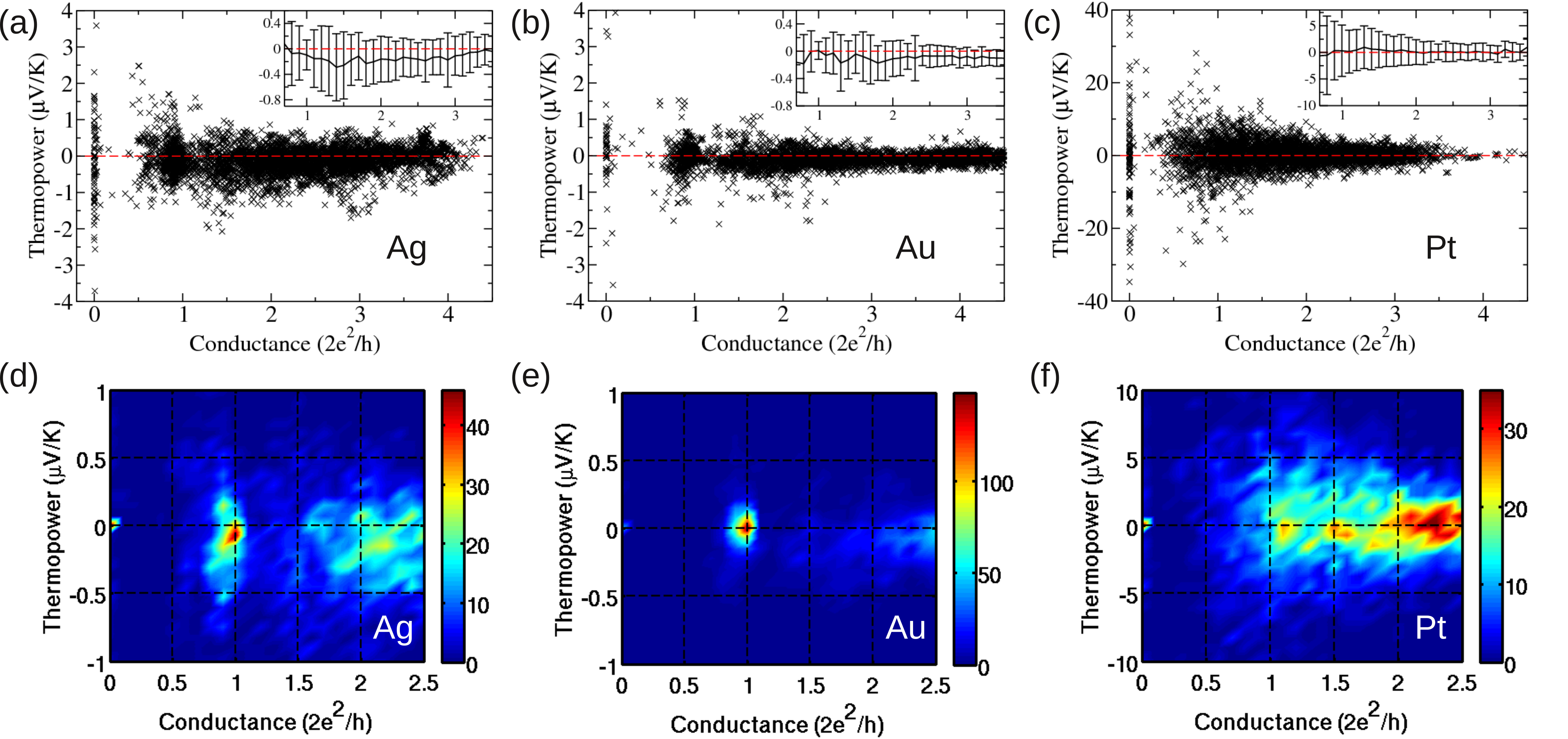}
\par\end{centering}

\caption{\label{fig:AgAuPt-SGhistos}(Color online) Scatter plots of the thermopower
as a function of the conductance for (a) Ag, (b) Au, and (c) Pt. The
main panels show all computed data points, while the insets display
the average of the thermopower together with its standard deviation.
Panels (d-f) are the corresponding density plots. For the insets of
plots (a-c) and the density plots (d-f), the bin size of the conductance
is $0.1G_{0}$. Those of the thermopower amount to $0.075$ $\mu$V/K
for (d,e) and $0.75$ $\mu$V/K for (f).}
\end{figure*}

The thermopower-conductance ($S$-$G$) plots are presented in Fig.~\ref{fig:AgAuPt-SGhistos}.%
 The main panels of Fig.~\ref{fig:AgAuPt-SGhistos}(a-c) show all the
computed data points. This representation provides an impression of
the overall scatter. In the insets the average value of the thermopower
$\langle S\rangle_{G}$ and the standard deviation $\sigma_{S}=\sqrt{\langle S^{2}\rangle_{G}-\langle S\rangle_{G}^{2}}$
are plotted. While $\left\langle S\right\rangle _{G}$ is very close
to zero for Pt, we find a trend towards negative values for Ag and
Au. From the theory of the free-electron gas in a hyperbolic constriction\cite{Torres:PRB1994}
a thermopower with a unique, negative sign is expected.\cite{Ludoph:PRB1999,Ludoph:PhD1999}
We cannot exclude such an origin for the trend, but our results with
thermopower values of both positive and negative sign demonstrate
that it is necessary to go beyond these simple free-electron-gas models
with idealized wire geometries to describe even the $s$-valent metallic
atomic contacts. Furthermore, within the standard deviation the results
for $\left\langle S\right\rangle _{G}$ are consistent with a vanishing
thermopower. From Fig.~\ref{fig:AgAuPt-SGhistos}(a-c) we also observe
that the variations of $S$ tend to increase for a decreasing minimum
cross-section of the contacts. Consistent with the sample opening
traces discussed above, the variations of the thermopower around zero
are predicted to be one order of magnitude larger for Pt than for
Ag and Au. 

With respect to $\left\langle S\right\rangle _{G}\approx0$ and increasing
$\sigma_{S}$ for decreasing $G$, the $S$-$G$ scatter plot for
Au closely resembles the experimental one of Ref.~\citenum{Ludoph:PRB1999}.
However, the absolute magnitude of the scatter of $S$ in our calculations
is too small by a factor of around 3, and we will discuss possible
reasons for this discrepancy below.\cite{Ludoph:PRB1999,Ludoph:PhD1999} 

\begin{figure}[!tb]
\begin{centering}
\includegraphics[width=1\columnwidth]{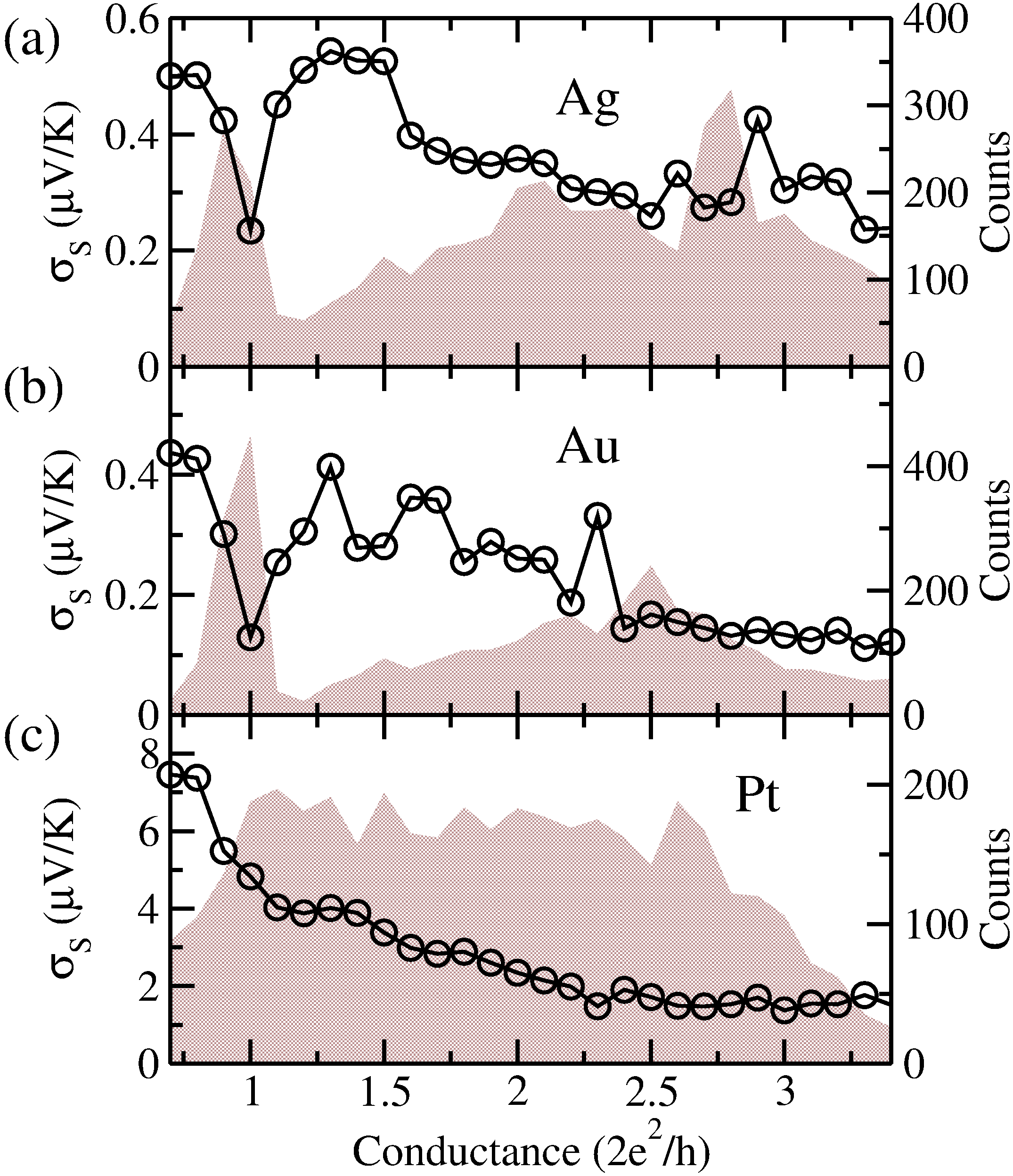}
\par\end{centering}

\caption{\label{fig:AgAuPt-stddevSG}(Color online) Standard deviation $\sigma_{S}$
of the thermopower as a function of the conductance for (a) Ag, (b)
Au, and (c) Pt. Shown in the background is the conductance histogram
for each metal, and the number of counts is indicated on the right
$y$-axis. In each case the bin size of the conductance is $0.1G_{0}$.}
\end{figure}

In order to reveal further differences between the mono- and multivalent
metals, we show the corresponding $S$-$G$ density plots in Fig.~\ref{fig:AgAuPt-SGhistos}(d-f).
For Ag and Au they exhibit a pronounced maximum at $(G,S)=(1G_{0},0)$,
while there is no particular feature for Pt at this position. The
peak for the $s$-valent metals seems plausible. In the conductance
histograms of Ag and Au pronounced maxima occur at $1G_{0}$.\cite{Ludoph:PRB2000,Yanson:PhD2001}
Our calculated conductance histograms, shown in Fig.~\ref{fig:AgAuPt-stddevSG}(a,b),
are consistent with this experimental finding.\cite{Dreher:PRB2005,Pauly:PRB2006}
Since the mean thermopower vanishes, maxima at $(1G_{0},0)$ are thus
expected in the density plots of Fig.~\ref{fig:AgAuPt-SGhistos}(d,e).
The experimental conductance histogram for Pt shows instead a rather
broad distribution between $1$ and $2G_{0}$ with a maximum at around
$1.5G_{0}$.\cite{Yanson:PhD2001,Smit:Nature2002,Kiguchi:PRL2008}
These features are not reproduced in detail in our calculations for
Pt in Fig.~\ref{fig:AgAuPt-stddevSG}(c), which shows a broad distribution
of frequently occurring conductances between $1$ and $3G_{0}$ without
a clear maximum. Possible reasons for the deviations, such as the
limited ensemble of considered junction geometries or the approximations
inherent to our method, have been discussed in Ref.~\citenum{Pauly:PRB2006}.
Still, from the broad distribution of conductance values no sharp
peak feature at a single conductance value is expected in the $S$-$G$
density plot, in agreement with Fig.~\ref{fig:AgAuPt-SGhistos}(f). 

While our discussion of Fig.~\ref{fig:AgAuPt-SGhistos}(d-f) considered
until now the conductance values, the sharpness of the peak feature
at $(1G_{0},0)$ with respect to the thermopower axis is somewhat
unexpected for Ag and Au. Further insight is obtained by plotting
the standard deviation $\sigma_{S}$ as a function of the conductance,
as displayed in Fig.~\ref{fig:AgAuPt-stddevSG}.%
 The calculations show indeed a suppression of $\sigma_{S}$ at $1G_{0}$
for Ag and Au, which explains the sharpness of the feature in the
corresponding $S$-$G$ density plots in Fig.~\ref{fig:AgAuPt-SGhistos}(d,e).
Suppressions of $\sigma_{S}$ at higher conductance values are absent
in Fig.~\ref{fig:AgAuPt-stddevSG}(a,b), while shallow minima were measured
for Au in that region.\cite{Ludoph:PRB1999} Consistent with the underestimation
of the scatter of $S$ in Fig.~\ref{fig:AgAuPt-SGhistos}(e), $\sigma_{S}$
for Au is smaller than in the experiments by a factor of 3. For the
multivalent metal Pt no particular features arise throughout the whole
range of conductance values considered in Fig.~\ref{fig:AgAuPt-stddevSG}(c),
especially there is no minimum of $\sigma_{S}$ at $1G_{0}$.

In order to describe the behavior of the thermopower fluctuations
in Fig.~\ref{fig:AgAuPt-stddevSG} in simple terms, we construct an {}``extended
single-level model'' (ESLM). We assume that the transmission is given
as \begin{equation}
\tau(E)=\sum_{n}\tau_{n}(E,\epsilon_{0}^{\left(n\right)},\Gamma_{L}^{\left(n\right)},\Gamma_{R}^{\left(n\right)})\label{eq:tau_ESLM}\end{equation}
with \begin{equation}
\tau_{n}(E,\epsilon_{0}^{\left(n\right)},\Gamma_{L}^{\left(n\right)},\Gamma_{R}^{\left(n\right)})=\frac{\Gamma_{L}^{\left(n\right)}\Gamma_{R}^{\left(n\right)}}{\left(E-\epsilon_{0}^{\left(n\right)}\right)^{2}+\frac{\left(\Gamma_{L}^{\left(n\right)}+\Gamma_{R}^{\left(n\right)}\right)^{2}}{4}}.\label{eq:taun_ESLM}\end{equation}
The function $\tau_{n}(E,\epsilon_{0}^{\left(n\right)},\Gamma_{L}^{\left(n\right)},\Gamma_{R}^{\left(n\right)})$
describes a resonance in the energy-dependent transmission probability
of the $n$th eigenchannel, which is located close to the Fermi energy.
The resonances arise from the atomic level structure at the narrowest
part of the contact and quantum interference effects related to the
disorder in the CW. Hence, the {}``level'' $\epsilon_{0}^{\left(n\right)}$
specifies the position of the resonance for the $n$th eigenchannel,
and $\Gamma_{L}^{\left(n\right)},\Gamma_{R}^{\left(n\right)}$ determine
the width and height of the resonance. For an atomic contact described
by $N$ eigenchannels, we can obtain from Eqs.~(\ref{eq:tau_ESLM}) and (\ref{eq:taun_ESLM})
the conductance and thermopower using the low-temperature expressions
of Eqs.~(\ref{eq:G}) and (\ref{eq:S}) for a given realization of parameters $\epsilon_{0}^{\left(n\right)},\Gamma_{L}^{\left(n\right)},\Gamma_{R}^{\left(n\right)}$
with $n=1,\ldots,N$. As explained in detail in Appendices \ref{sec:AvChanT}
and \ref{sec:ESLM}, we identify the $\tau_{n}(E_{F},\epsilon_{0}^{\left(n\right)},\Gamma_{L}^{\left(n\right)},\Gamma_{R}^{\left(n\right)})$
with respective eigenchannel transmission probabilities $\tau_{n}$
in order to distinguish between mono- and multivalent metals.

As the simplest model for a monovalent metal we assume that the transmission
channels open one by one, i.e.\ that $\tau_{1}=\ldots=\tau_{n-1}=1$
and $\tau_{n}=G/G_{0}-(n-1)$ for $n-1\leq G/G_{0}\leq n$. In this
case (see Appendix \ref{sec:ESLM}) we find that \begin{equation}
\sigma_{S}=\alpha\sqrt{n-G/G_{0}}/n\label{eq:sigmaSmono}\end{equation}
for $G\rightarrow nG_{0}$ in the conductance interval $(n-1)G_{0}\leq G\leq nG_{0}$,
and $\alpha$ is a constant prefactor of dimension V/K. This relation
explains the minimum of $\sigma_{S}$ at $1G_{0}$ as visible in Fig.~\ref{fig:AgAuPt-stddevSG}(a,b).
Minima at higher conductance values $nG_{0}$ with $n>1$ are absent
for Ag and Au, since the assumption that channels open one by one,
also known as the {}``saturation of channel transmission'',\cite{Ludoph:PRL1999,vandenBrom:PRL1999,Ludoph:PRB1999,Ludoph:PRB2000}
is accurately obeyed in our calculations only for the first channel,
i.e.\ for conductances below $1G_{0}$ (see Appendix \ref{sec:AvChanT}
and Refs.~\citenum{Dreher:PRB2005,Pauly:PRB2006}). The presence
of minima at integer conductance values above $1G_{0}$ is hence hard
to understand from our simulations. Frequently occurring junction
geometries with several atomic chains in parallel\cite{Ohnishi:Nature1998}
may provide an explanation, and investigations of junctions with larger
initial diameters could help to resolve the puzzle.

In contrast to the monovalent metals the ESLM of a multivalent metal,
as presented in Appendix \ref{sec:ESLM}, predicts that minima of
$\sigma_{S}$ are absent, or at least strongly reduced, due to the
contribution of several partially open transmission channels throughout
the whole range of conductance values. This is consistent with our
results from the atomistic simulations in Fig.~\ref{fig:AgAuPt-stddevSG}(c).

The behavior of the thermopower has been discussed in terms of a different
and very detailed model in Refs.~\citenum{Ludoph:PRB1999,Ludoph:PRB2000},
which is applicable to the whole range of conductance values. However,
apart from the prefactor, at $G$ below but close to $nG_{0}$ their
expression for $\sigma_{S}$ in monovalent metals is identical to
our Eq.~(\ref{eq:sigmaSmono}). As a common feature, both our and their
approach explain the suppression of the thermopower fluctuations by
disorder-related quantum interference effects (see also Appendix \ref{sec:ESLM}).
In their model the thermopower fluctuations arise from the interference
of directly transmitted electrons and those backscattered elastically
in the diffusive regions in the vicinity of the narrowest part of
the contact. The backscattering contributions are considered to lowest
order, and channel transmissions in the ballistic central region are
assumed to be energy-independent and fixed for a certain $G$. In
contrast, in our atomistic simulations we determine the energy-dependent
transmission of the disordered CW, take scattering effects into account
to all orders, and different realizations of the set of $\tau_{n}$
may contribute for a given value of the conductance.

In the experiments both local and nonlocal interference contributions
will be present and backscattering from defects up to the coherence
length of some 100 nm away from the constriction has been reported
at the low temperatures relevant here.\cite{Untiedt:PRB2000} Such
highly nonlocal interference contributions are clearly not described
by our short CWs. This may result in an underestimation of the energy
dependence of the transmission, causing the factor of 3 discrepancy
with respect to the experimental values in Figs.~\ref{fig:AgAuPt-SGhistos}(b)
and \ref{fig:AgAuPt-stddevSG}(b) mentioned above.\cite{Ludoph:PRB1999}
Another reason for the discrepancy may be the large ensemble of experimentally
realized contact configurations, not fully taken into account in our
calculations.

While measurements of the thermopower and its fluctuations are not
yet available for Pt, Ludoph and van Ruitenbeek have shown within
their model that $\sigma_{S}$ is proportional to the standard deviation
$\sigma_{GV}$ of the voltage-dependent conductance.\cite{Ludoph:PRB2000}
The comparison of their results for $\sigma_{GV}$ for the transition
metals niobium and iron and the $s$-valent metals Cu, Ag, and Au
shows larger fluctuations by a factor of two for the former. Our prediction
of a factor of 10 larger fluctuations for the thermopower of Pt as
compared with the $s$-valent metals may be considered as an upper
bound, and a reduction might arise from the finite averaging times
in the measurements, for instance.

As another important aspect we have discussed only the elastic electronic
contribution to the thermopower. Inelastic effects due to the electron-phonon
coupling may lead to modifications, which we expect to be small, however,
for the experimental conditions considered here. Thus, due to its
weakness the electron-phonon coupling can typically be treated perturbatively
for metallic atomic contacts,\cite{Agrait:PRL2002,Frederiksen:PRL2004,Viljas:PRB2005,Hsu:PRB2011}
and the phonon-drag contributions should be suppressed because of
the small contact diameter, the low measurement temperatures, and
small applied thermal gradients.\cite{Bogacheck:JETPL1985,Shklyarevskii:PRL1986,Ludoph:PRB1999}
Our results, with the reasonable agreement between experiment and
theory, show that the elastic contribution is sufficient to describe
the main experimental features.

\section{Conclusions and outlook\label{sec:Conclusions}}

Using molecular dynamics simulations of up to 50 stretching events,
we have analyzed the thermopower of atomic contacts of Ag, Au, and
Pt. For Ag and Au its behavior agrees well with previously reported
measurements.\cite{Ludoph:PRB1999} On a quantitative level, however,
the experimental scatter of the thermopower values is underestimated
by a factor of around 3.\cite{Ludoph:PRB1999,Ludoph:PhD1999} The
thermopower-conductance plots show the thermopower to be zero on average
within the standard deviation for the three metals studied. Furthermore
our calculations predict its variations around the mean to increase
for the narrowest constrictions and to be one order of magnitude larger
for the multivalent Pt as compared to the $s$-valent metals Ag and
Au. At a conductance of one quantum of conductance we find, in agreement
with the experiment, a suppression of the fluctuations of the thermopower
for Ag and Au. For the multivalent metal Pt possible minima of the
fluctuations of the thermopower should be shifted to higher conductance
values, but they are predicted to be absent, or at least strongly
reduced, due to the influence of several partially open transmission
eigenchannels.

Our calculations indicate that the essential characteristics of the
thermopower of metallic atomic-size contacts can be understood based
on the elastic electronic contribution combined with effects of rather
local disorder at the narrowest part of the atomic contacts. Quantitative
differences between our simulations and the experiment may arise from
the larger variability of contact configurations in the experiments
and distant scatterers, not taken into account in the calculations
due to the limited system size.\cite{Ludoph:PRB2000} The quantification
of effects due to the electron-phonon coupling, which are expected
to be small for metallic single-atom contacts,\cite{Bogacheck:JETPL1985,Shklyarevskii:PRL1986,Ludoph:PRB1999,Hsu:PRB2011}
constitutes a challenging task for the future.
\begin{acknowledgments}
We are grateful to J.~M.~van Ruitenbeek, E.~Scheer, and G.~Sch\"on
for stimulating discussions. M.B.\ was funded by the CFN and the
DFG SPP 1243, as well as by the Baden-W\"urttemberg Stiftung within
the Network of Excellence \textquotedblleft{}Functional Nanostructures''.
J.K.V.\ acknowledges financial support from the Academy of Finland,
M.D. and P.N. from the SFB 767 and the NIC, J.C.C.\ from the Spanish
MICINN (Contract No.\ FIS2008-04209), and F.P.\ from the Young Investigator
Group.
\end{acknowledgments}
\appendix

\section{Further junction stretching events\label{sec:stretchAdd}}

\begin{figure}[!t]
\begin{centering}
\includegraphics[width=1\columnwidth]{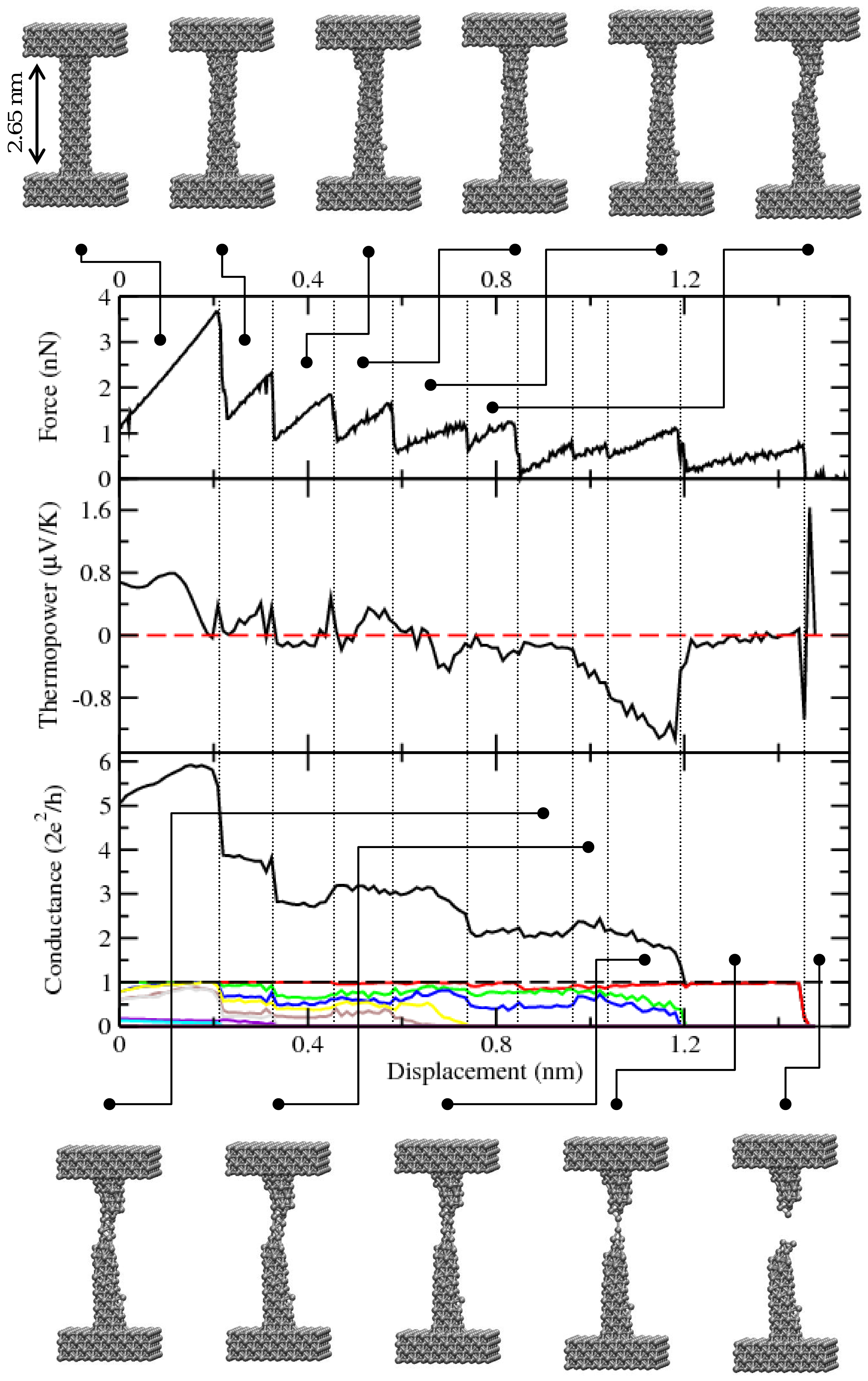}
\par\end{centering}

\caption{\label{fig:Ag2-v2-contact}(Color online) The same as Fig.~\ref{fig:Ag2-contact}
for another Ag contact forming a dimer before rupture.}
\end{figure}

\begin{figure}[!t]
\begin{centering}
\includegraphics[width=1\columnwidth]{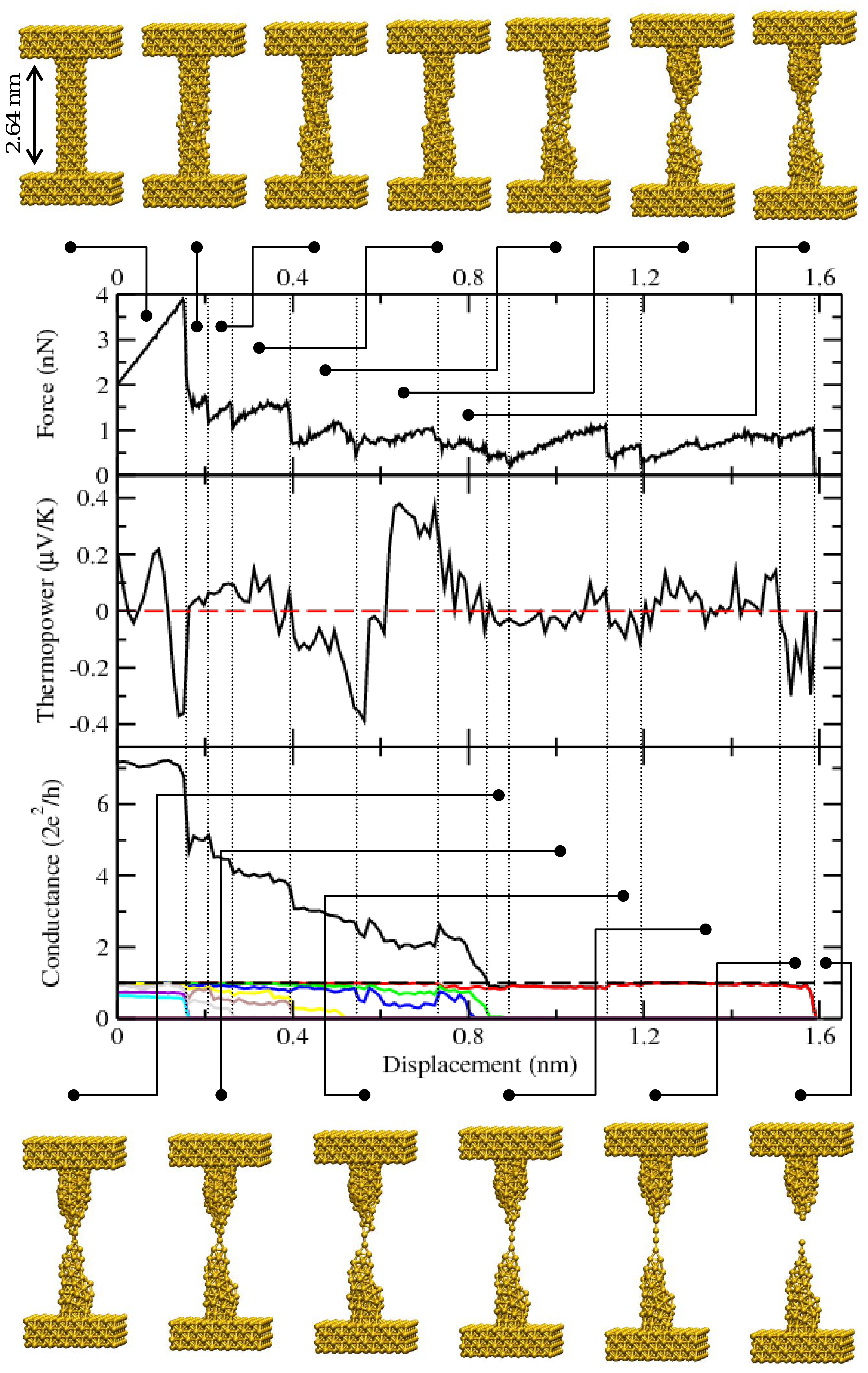}
\par\end{centering}

\caption{\label{fig:Au-c4-contact}(Color online) The same as Fig.~\ref{fig:Ag2-contact},
but for a Au contact forming a chain of four atoms before rupture.}
\end{figure}

\begin{figure}[!t]
\begin{centering}
\includegraphics[width=1\columnwidth]{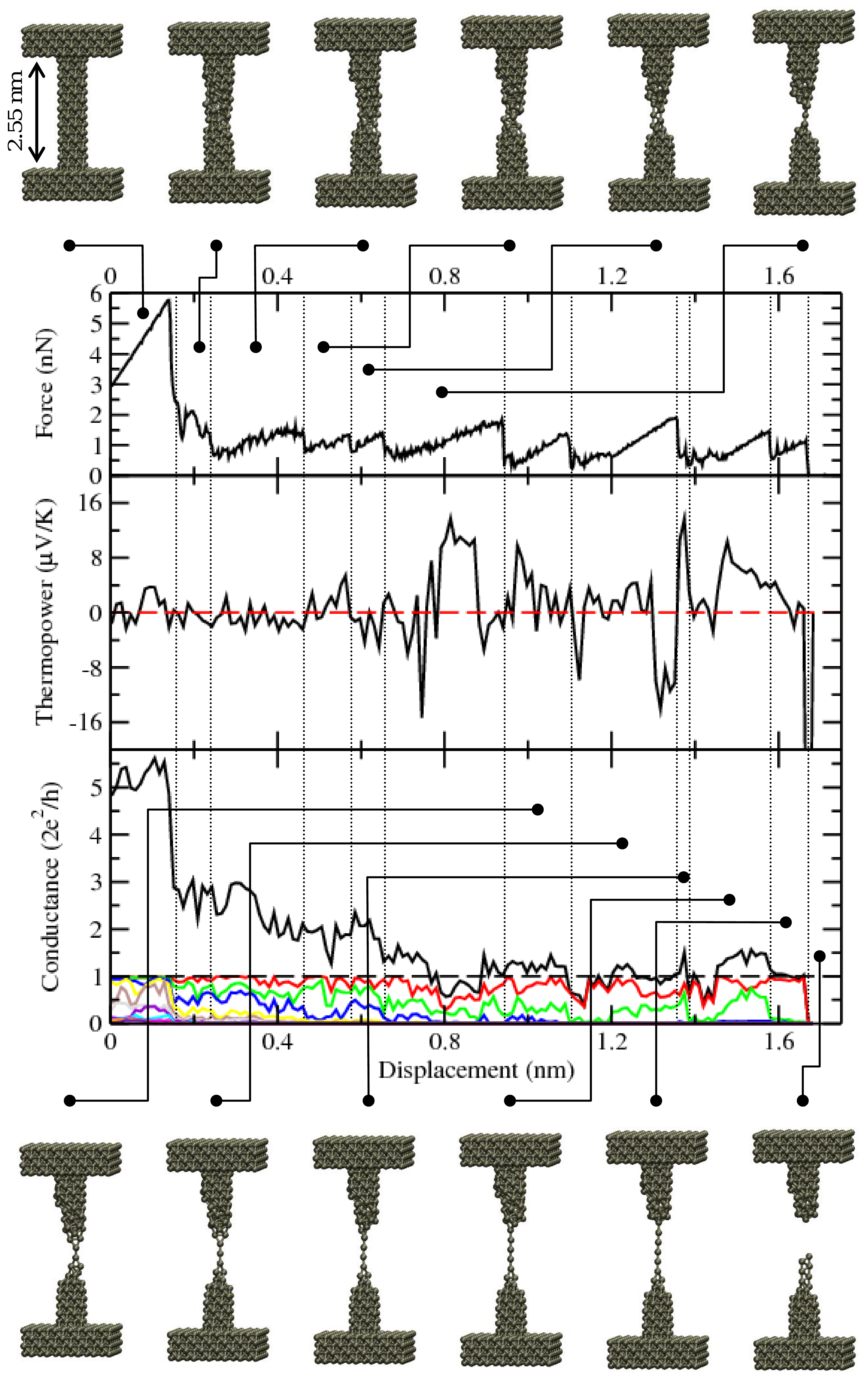}
\par\end{centering}

\caption{\label{fig:Pt-c6-contact}(Color online) The same as Fig.~\ref{fig:Ag2-contact},
but for a Pt contact forming a chain of six atoms before rupture.}
\end{figure}

Here, we provide further examples for the evolution of junction properties
upon stretching for each of the three metals studied. For Au and Pt
we have selected junctions forming chains of several atoms in length
before rupture. Animations showing the stretching processes displayed
in Figs.~\ref{fig:Ag2-contact} to \ref{fig:Pt2-contact} and
Figs.~\ref{fig:Ag2-v2-contact} to \ref{fig:Pt-c6-contact}
can be found as supplemental material on the web.\cite{EPAPS}

In Fig.~\ref{fig:Ag2-v2-contact} we show an Ag contact.%
 After the initial stage, the conductance evolves from a value of
around $4G_{0}$ via various plateaus to around $2G_{0}$ until a
rather stable dimer is formed. For the dimer contact the conductance
is pinned closely to $1G_{0}$ and, at the same time, the thermopower
and its fluctuations are suppressed to zero. This particular example
agrees well with the observation of the small $\left\langle S\right\rangle _{G}$
and minimum of $\sigma_{S}$ for Ag, as shown in
Figs.~\ref{fig:AgAuPt-SGhistos} and \ref{fig:AgAuPt-stddevSG},
when the conductance arises from a single completely open conduction
channel.

For Au we display in Fig.~\ref{fig:Au-c4-contact} a contact forming a
four-atom chain before breaking.%
 The chain is increasing successively in length, starting from a single
atom for electrode displacements between $0.84$ and $1.12$ nm. Two
of them are present from $1.12$ to $1.2$ nm, three from $1.2$ to
$1.51$ nm, and four from $1.51$ to $1.59$ nm. As soon as the single-atom
contact has formed, only a single channel contributes to the conductance.
For the one- to three-atom chain configurations the conductance is
very close to $1G_{0}$ and the thermopower varies around zero with
a relatively small amplitude. When the four-atom chain has formed,
the conductance is somewhat suppressed and the thermopower increases
in absolute magnitude.

In Fig.~\ref{fig:Pt-c6-contact} we show a Pt contact forming a chain
of six atoms before rupture.%
 As in Fig.~\ref{fig:Pt2-contact} the conductance and thermopower during
the stretching show much larger variations than for Ag and Au, and
also the typical values of the thermopower are around one order of
magnitude larger than for the $s$-valent metals. After the single-atom
contact has formed at around $0.66$ nm, more atoms are incorporated
into the chain at the elongations marked with dotted lines (i.e.\
at $0.95$, $1.11$, $1.36$, $1.38$, and $1.59$ nm), until the
final length of six atoms is reached. In this atomic-chain regime
we generally observe two to three partially open transmission channels,
often leading to conductances exceeding $1G_{0}$. In addition, for
these narrowest contacts the thermopower shows its largest fluctuations
with the average remaining close to zero.

\section{Average eigenchannel transmission probabilities\label{sec:AvChanT}}

For the monovalent metals transmission eigenchannels are typically
assumed to open one by one, while for multivalent metals there are
several channels contributing already at the lowest conductances.\cite{Ludoph:PRL1999,vandenBrom:PRL1999,Ludoph:PRB2000}
We can inspect the validity of these assertions by examining the average
eigenchannel transmission probabilities obtained from our MD transport
simulations as a function of the junction conductance.\cite{Dreher:PRB2005,Pauly:PRB2006} 

\begin{figure*}[!]
\begin{centering}
\includegraphics[width=2\columnwidth]{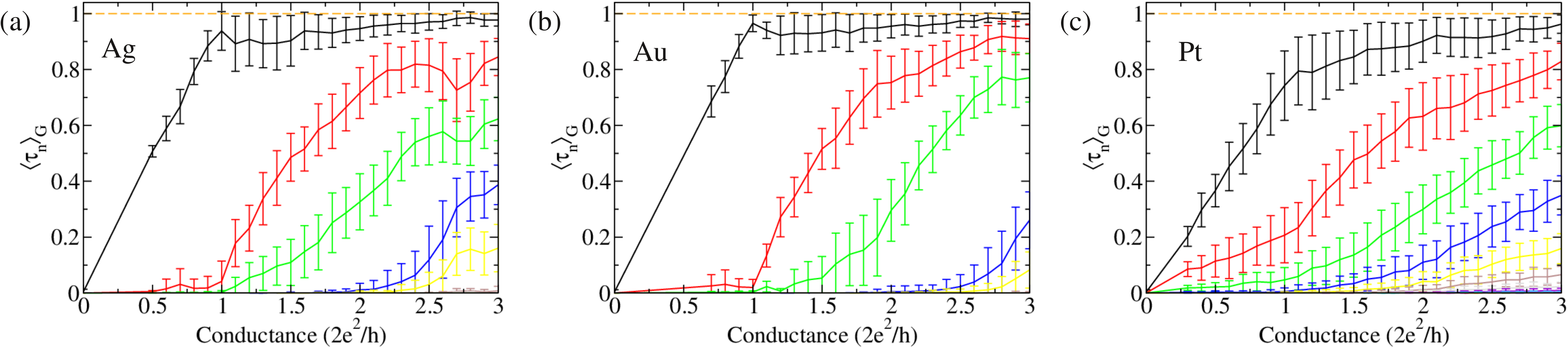}
\par\end{centering}

\caption{\label{fig:AvChanT}(Color online) Average eigenchannel transmission
probabilities $\left\langle \tau_{n}\right\rangle _{G}$ as a function
of the conductance for (a) Ag, (b) Au, and (c) Pt, as determined from
the atomistic simulations presented in the manuscript. Error bars
show the standard deviation and the dashed horizontal lines indicate
unit average transmission. See also Refs.~\citenum{Dreher:PRB2005,Pauly:PRB2006}
for further explanations.}
\end{figure*}

Using the results of the present study, we display the average eigenchannel
transmissions in Fig.~\ref{fig:AvChanT}.%
 We do indeed observe that for the monovalent metals there is essentially
only a single open channel for $0\leq G\leq1G_{0}$. However, the
first channel is not fully open at $1G_{0}$, i.e.\ does not reach
a transmission of one. For $G\geq1G_{0}$ the saturation of channel
transmission still seems to be a reasonable approximation for Au.
However, the results for Ag rather suggest an interpretation in terms
of a simultaneous opening of two $\pi$-like channels, whose degeneracy
is lifted in the disordered junction geometries. For Pt there are
significant contributions from at least two partially open channels
throughout the whole range of conductance values.

\section{Fluctuations of the thermopower within the extended single-level
model\label{sec:ESLM}}

\subsection{Extended single-level model}

We consider a model for the charge transport through metallic atomic
contacts in order to explain the results of our atomistic simulations
in simple terms. We assume that the energy-dependent transmission
$\tau(E)$ arises from those of independent single-level transmissions,
as described by Eqs.~(\ref{eq:tau_ESLM}) and (\ref{eq:taun_ESLM}). Setting the Fermi
energy to zero, $E_{F}=0,$ we can determine the conductance and thermopower
in the limit of low temperatures as\begin{equation}\begin{split}\label{eq:G_ESLM}
G &= G_{0}\left.\tau(E)\right|_{E=E_{F}} \\
&= G_{0}\sum_{n}\frac{\Gamma_{L}^{\left(n\right)}\Gamma_{R}^{\left(n\right)}}{\left(\epsilon_{0}^{\left(n\right)}\right)^{2}+\left(\Gamma_{L}^{\left(n\right)}+\Gamma_{R}^{\left(n\right)}\right)^{2}/4},
\end{split}\end{equation}
\begin{equation}\begin{split}\label{eq:S_ESLM}
S &= -S_{0}\left.\frac{\partial_{E}\tau(E)}{\tau(E)}\right|_{E=E_{F}} \\
&= -S_{0}\sum_{n}\frac{2\epsilon_{0}^{\left(n\right)}\Gamma_{L}^{\left(n\right)}\Gamma_{R}^{\left(n\right)}}{\left[\left(\epsilon_{0}^{\left(n\right)}\right)^{2}+\left(\Gamma_{L}^{\left(n\right)}+\Gamma_{R}^{\left(n\right)}\right)^{2}/4\right]^{2}}/G
\end{split}\end{equation}with $G_{0}=2e^{2}/h$ and $S_{0}=\pi^{2}k_{B}^{2}T/3e$.

During the stretching of a contact the atomic positions change, causing
related variations in the electronic structure and in the quantum
interference pattern. Therefore, we consider the parameters $\epsilon_{0}^{\left(n\right)},\Gamma_{L}^{\left(n\right)},\Gamma_{R}^{\left(n\right)}$
to be independent random variables distributed with probability densities
$P_{\epsilon}(\epsilon_{0}^{\left(n\right)})$, $P_{\Gamma}(\Gamma_{L}^{\left(n\right)})$,
$P_{\Gamma}(\Gamma_{R}^{\left(n\right)})$, respectively. For the
discussion of the fluctuations of the thermopower, we determine the
standard deviation \begin{equation}
\sigma_{S}(G)=\sqrt{\left\langle S^{2}\right\rangle _{G}-\left\langle S\right\rangle _{G}^{2}},\end{equation}
where $\left\langle \;\right\rangle _{G}$ means a conditional average
over all configurations yielding the conductance $G$. For a system
of $N$ levels the configurations are labeled by $x=(\epsilon_{0}^{\left(1\right)},\Gamma_{L}^{\left(1\right)},\Gamma_{R}^{\left(1\right)},\ldots,\epsilon_{0}^{\left(N\right)},\Gamma_{L}^{\left(N\right)},\Gamma_{R}^{\left(N\right)})$.
By introducing the probability density \begin{equation}
p(x)=\prod_{n=1}^{N}P_{\epsilon}(\epsilon_{0}^{\left(n\right)})P_{\Gamma}(\Gamma_{L}^{\left(n\right)})P_{\Gamma}(\Gamma_{R}^{\left(n\right)}),\end{equation}
we can express such an average as \begin{equation}
\left\langle S\right\rangle _{G}=\int dxS(x)p(x|G)\end{equation}
with the conditional probability density $p(x|G)=\delta(G-G(x))p(x)/\int dx\delta(G-G(x))p(x)$.
Here, $G(x)$ and $S(x)$ are determined from Eqs.~(\ref{eq:G_ESLM}) and (\ref{eq:S_ESLM}),
respectively.

In order to describe material-specific properties and to understand
the behavior of $\sigma_{S}$ in Fig.~\ref{fig:AgAuPt-stddevSG}, we need
to know how the individual resonances $\tau_{n}(E,\epsilon_{0}^{\left(n\right)},\Gamma_{L}^{\left(n\right)},\Gamma_{R}^{\left(n\right)})$
of the eigenchannels in Eq.~(\ref{eq:taun_ESLM}) contribute to the transmission
for a given conductance. For this purpose, we identify $\tau_{n}(E_{F},\epsilon_{0}^{\left(n\right)},\Gamma_{L}^{\left(n\right)},\Gamma_{R}^{\left(n\right)})$
with the measurable eigenchannel transmission probability $\tau_{n}$
to complete the ESLM. This identification procedure will be explained
further in the next subsections.

\subsection{Monovalent metals}

As visible from Fig.~\ref{fig:AvChanT}, the saturation of channel transmission
is well fulfilled in our simulations only for $G\leq1G{}_{0}$. However,
to compare to the literature\cite{Ludoph:PRB1999,vandenBrom:PRL1999}
and since we are mainly interested in conductance values around $1G_{0}$,
we adopt the following ideal model to describe the monovalent metals.
For a given $G$ with $n-1\leq G/G_{0}\leq n$, we determine the transmission
as \begin{equation}
\tau(E)=n-1+\tau_{n}(E,\epsilon_{0}^{\left(n\right)},\Gamma_{L}^{\left(n\right)},\Gamma_{R}^{\left(n\right)}).\label{eq:Tmodel_AgAu}\end{equation}
Hence, $\tau_{1}=\ldots=\tau_{n-1}=1$, $\tau_{n}=G/G_{0}-\left(n-1\right)$
for $n-1\leq G/G_{0}\leq n$. This behavior of the transmission probabilities
$\tau_{n}=\tau_{n}(E_{F},\epsilon_{0}^{\left(n\right)},\Gamma_{L}^{\left(n\right)},\Gamma_{R}^{\left(n\right)})$
as a function of the conductance is visualized in Fig.~\ref{fig:sSGsimul}(a).

In Fig.~\ref{fig:AgAuPt-stddevSG} it is seen that $\sigma_{S}$ is suppressed
at $G\approx1G_{0}$ for Ag and Au. In order to analyze this, we consider
a conductance $G$ close to but smaller than $nG_{0}$ in the ESLM.
For $(n-1)\leq G/G_{0}\leq n$ we obtain\begin{equation}
G/G_{0}=n-1+\frac{\Gamma_{L}^{\left(n\right)}\Gamma_{R}^{\left(n\right)}}{\left(\epsilon_{0}^{\left(n\right)}\right)^{2}+\left(\Gamma_{L}^{\left(n\right)}+\Gamma_{R}^{\left(n\right)}\right)^{2}/4},\label{eq:Gmono}\end{equation}
\begin{equation}
S/S_{0}=-\frac{2\epsilon_{0}^{\left(n\right)}}{\Gamma_{L}^{\left(n\right)}\Gamma_{R}^{\left(n\right)}}\frac{\left(G/G_{0}+1-n\right)^{2}}{G/G_{0}}.\label{eq:Smono}\end{equation}
Since $\tau_{n}\approx1$ for $G\approx nG_{0}$, Eq.~(\ref{eq:Gmono})
requires $\Gamma_{L}^{\left(n\right)}\approx\Gamma_{R}^{\left(n\right)}$
and $\epsilon_{0}^{\left(n\right)}\approx0$. Therefore, we consider
the symmetric junction model with $\Gamma=\Gamma_{L}^{\left(n\right)}=\Gamma_{R}^{\left(n\right)}\geq0$
and $\epsilon_{0}=\epsilon_{0}^{\left(n\right)}$. Since we are dealing
with atomic contacts, we can expect $P_{\Gamma}(\Gamma)$ to be centered
at a nonvanishing, positive value, while we assume $P_{\epsilon}(\epsilon_{0})$
to be symmetric and maximal at $\epsilon_{0}=0$. With these assumptions,
it follows that \begin{equation}
\sigma_{S}^{2}=\left\langle S^{2}\right\rangle _{G}=4S_{0}^{2}\left\langle \frac{1}{\Gamma^{2}}\right\rangle _{G}\frac{\left(G/G_{0}+1-n\right)^{3}}{\left(G/G_{0}\right)^{2}}(n-G/G_{0}),\label{eq:sigma2Smono}\end{equation}
where $\left\langle 1/\Gamma^{2}\right\rangle _{G}$ approaches a
fixed value in the limit $G\rightarrow nG_{0}$. From the expression,
we hence obtain Eq.~(\ref{eq:sigmaSmono}).

\subsection{Multivalent metals}

\begin{figure*}[!t]
\begin{centering}
\includegraphics[width=1.5\columnwidth]{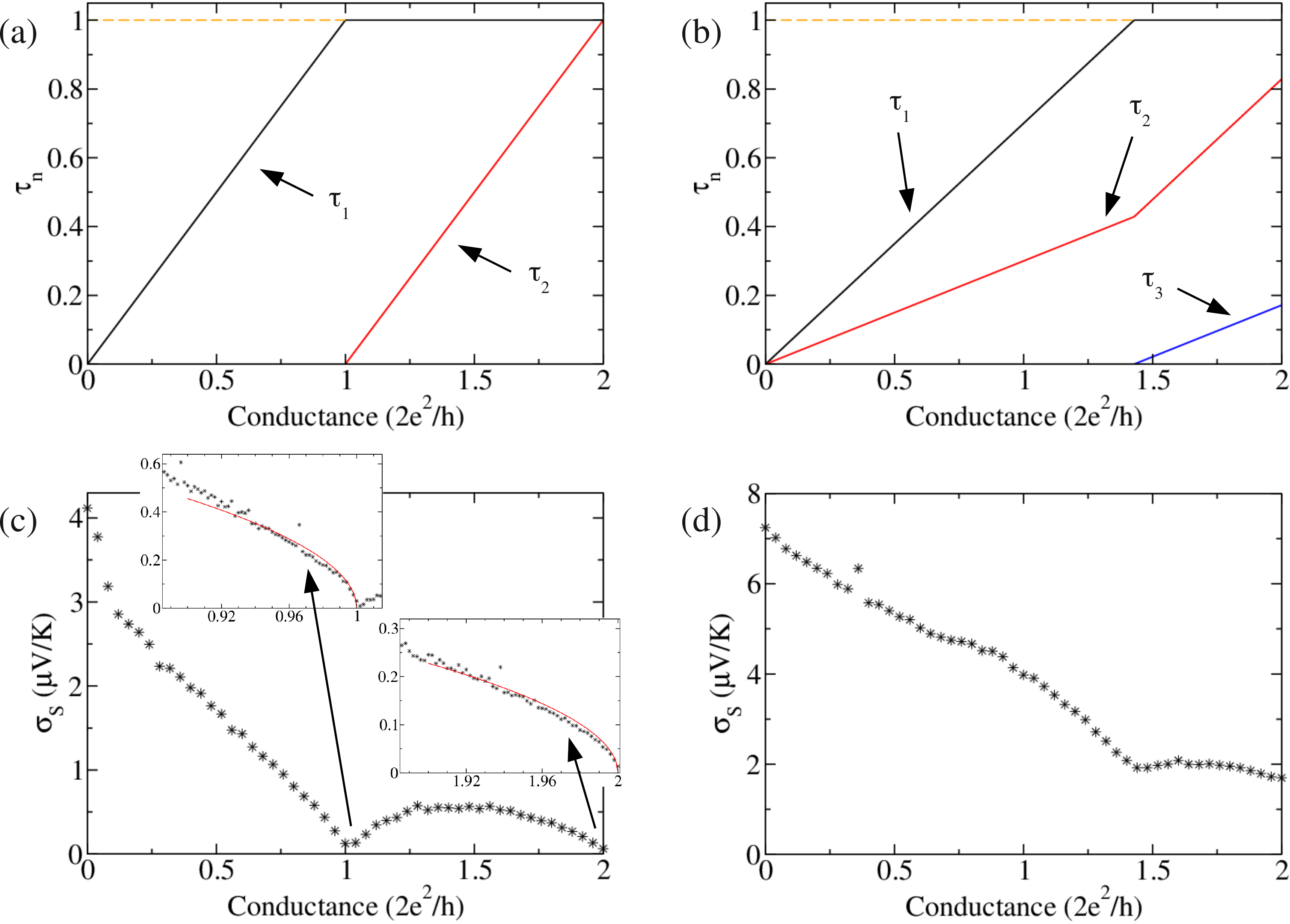}
\par\end{centering}

\caption{\label{fig:sSGsimul}(Color online) Behavior of the eigenchannel transmission
probabilities $\tau_{n}$ and the standard deviation of the thermopower
$\sigma_{S}$ as a function of the conductance for (a,c) a monovalent
metal and (b,d) a multivalent metal. For (c) and (d) we used $\gamma=0.6$
eV and $0.1$ eV, respectively, and the bin size was set to $\Delta G=0.04G_{0}$
in the main panels. In the insets of panel (c) we have chosen a smaller
bin size of $\Delta G=0.002G_{0}$ to resolve the selected region
more clearly and show the function $\sigma_{S}=\alpha\sqrt{n-G/G_{0}}/n$
for $G\lesssim nG_{0}$ and $n=1,2$ as a red line, assuming $\alpha=1.44$
$\mu$V/K.}
\end{figure*}
Based on the results in Fig.~\ref{fig:AvChanT}(c), we want to construct
an ESLM for the multivalent Pt. To simplify the situation, we consider
only two partially open channels and write\begin{widetext} \begin{equation}
\tau(E)=\begin{cases}
\tau_{1}(E,\epsilon_{0}^{\left(1\right)},\Gamma_{L}^{\left(1\right)},\Gamma_{R}^{\left(1\right)})+\tau_{2}(E,\epsilon_{0}^{\left(2\right)},\Gamma_{L}^{\left(2\right)},\Gamma_{R}^{\left(2\right)}) & \mathrm{if}\;0\leq G/G_{0}\leq1/a_{1}\\
1+\tau_{2}(E,\epsilon_{0}^{\left(2\right)},\Gamma_{L}^{\left(2\right)},\Gamma_{R}^{\left(2\right)})+\tau_{3}(E,\epsilon_{0}^{\left(3\right)},\Gamma_{L}^{\left(3\right)},\Gamma_{R}^{\left(3\right)}) & \mathrm{if}\;1/a_{1}<G/G_{0}\leq\left(2a_{1}-a_{2}\right)/a_{1}^{2}\end{cases},\label{eq:Tmodel_Pt}\end{equation}
\end{widetext}with $\tau_{1}=a_{1}G/G_{0}$, $\tau_{2}=a_{2}G/G_{0}$
for $0\leq G/G_{0}\leq1/a_{1}$, and $\tau_{1}=1$, $\tau_{2}=a_{2}/a_{1}+a_{1}(G/G_{0}-1/a_{1})$,
$\tau_{3}=a_{2}(G/G_{0}-1/a_{1})$ for $1/a_{1}<G/G_{0}\leq\left(2a_{1}-a_{2}\right)/a_{1}^{2}$.
The model requires $a_{1}+a_{2}=1$, and $a_{1}=0.7$ seems to be
a reasonable choice for Pt. The upper bound for the conductance in
Eq.~(\ref{eq:Tmodel_Pt}) considers in each case the full opening of a
channel, and the model can easily be extended to include further partially
open channels or describe larger values of $G$. The behavior of the
eigenchannel transmission probabilities as a function of the conductance
is visualized in Fig.~\ref{fig:sSGsimul}(b).

We note that in the conductance range $0\leq G/G_{0}\leq1/a_{1}$,
for a given realization $\epsilon_{0}^{\left(1\right)},\Gamma_{L}^{\left(1\right)},\Gamma_{R}^{\left(1\right)}$
yielding $\tau_{1}=\tau_{1}(E_{F},\epsilon_{0}^{\left(1\right)},\Gamma_{L}^{\left(1\right)},\Gamma_{R}^{\left(1\right)})$,
we guess values for $\epsilon_{0}^{\left(2\right)},\Gamma_{L}^{\left(2\right)},\Gamma_{R}^{\left(2\right)}$
until $(1-p)a_{2}\tau_{1}/a_{1}\leq\tau_{2}\leq(1+p)a_{2}\tau_{1}/a_{1}$
with $\tau_{2}=\tau_{2}(E_{F},\epsilon_{0}^{\left(2\right)},\Gamma_{L}^{\left(2\right)},\Gamma_{R}^{\left(2\right)})$
and with a small tolerance parameter $p=0.05$. For conductance values
$G/G_{0}>1/a_{1}$ we proceed similarly.

\subsection{Statistical analysis of the fluctuations of the thermopower}

By treating the parameters $\epsilon_{0}^{\left(n\right)},\Gamma_{L}^{\left(n\right)},\Gamma_{R}^{\left(n\right)}$
as independent random numbers with $\left|\epsilon_{0}^{\left(n\right)}\right|\leq0.1$
eV and $0\leq\Gamma_{L}^{\left(n\right)},\Gamma_{R}^{\left(n\right)}\leq\gamma$,
we determine $\left(G,S\right)$ pairs from the prescriptions in
Eqs.~(\ref{eq:Tmodel_AgAu}) and (\ref{eq:Tmodel_Pt}).
We choose a bin size $\Delta G$ and perform the typical statistical
analysis. Specifically, we use $\left\langle S\right\rangle _{G}=\sum_{j}S_{j}/M$
where $S_{j}$ are all those thermopower values of the pairs with
$G-\Delta G/2\leq G_{j}\leq G+\Delta G/2$, and $M$ is the total
number of such pairs. Analogously, we obtain the standard deviation
of the thermopower as $\sigma_{S}=\sqrt{\sum_{j}(S_{j}-\left\langle S\right\rangle _{G})^{2}/M}$.

By assuming $\epsilon_{0}^{\left(n\right)}$ to vary symmetrically
around zero, we obtain $\langle S\rangle_{G}\approx0$ in the ESLM.
In this way we cannot describe the systematic deviations from zero
with a unique sign, predicted and measured for quantum point contacts
realized in a two-dimensional electron gas.\cite{Streda:JPhysCondMat1989,Molenkamp:PRL1990,Houten:SST1992}
However, our assumption is consistent with the experiments on atomic
contacts, where $\langle S\rangle_{G}$ was found to scatter around
zero largely within the $\sigma_{S}$.\cite{Ludoph:PhD1999,Ludoph:PRB1999}

From the numerical analysis we obtain the $\sigma_{S}(G)$ curves
shown in Fig.~\ref{fig:sSGsimul}(c,d). In order to obtain standard deviations
comparable to the atomistic simulations we have set the maximum of
the couplings to $\gamma=0.6$ eV for the monovalent metal. Because
of the sharper resonances due to $d$ states for Pt (compare the Figs.~3,
7, and 9 in Ref.~\citenum{Pauly:PRB2006}) a smaller value $\gamma=0.1$
eV is needed for the model of the multivalent metal. Based on Eq.~(\ref{eq:sigma2Smono})
we observe that a smaller $\gamma$ should generally increase $\sigma_{S}$.
To obtain ESLM results for $\sigma_{S}$ of the same size as in the
experiments on Au,\cite{Ludoph:PRB1999} $\gamma$ should correspond
to around 0.2 eV. Since we would expect line widths $\Gamma_{L}^{\left(n\right)},\Gamma_{R}^{\left(n\right)}$
around $1$ eV for atomic levels, the small values of even the \emph{maximum}
couplings are consistent with the interpretation that the resonances
$\tau_{n}(E,\epsilon_{0}^{\left(n\right)},\Gamma_{L}^{\left(n\right)},\Gamma_{R}^{\left(n\right)})$
are not simply due to atomic orbitals in the narrowest part of the
contact, but that they arise essentially from disorder-related quantum
interference effects.

A clear suppression of the fluctuations of the thermopower is visible
in the regions $G\approx nG_{0}$ in Fig.~\ref{fig:sSGsimul}(c) for the
model of a monovalent metal. The insets of that panel illustrate that
$\sigma_{S}=\alpha\sqrt{n-G/G_{0}}/n$ at $G\approx nG_{0}$, as expected
from Eq.~(\ref{eq:sigmaSmono}). The suppression at $1G_{0}$ is consistent
with the results of our atomistic simulations and the experiments.\cite{Ludoph:PRB1999}
For Ag and Au the dip at $1G_{0}$ in Fig.~\ref{fig:AgAuPt-stddevSG}(a,b)
is reduced in depth as a result of the incomplete opening of the first
conductance channel, the small contribution of further partially open
conduction channels {[}see Fig.~\ref{fig:AvChanT}(a,b){]}, and the limited
statistics with the related large $\Delta G=0.1G_{0}$. These effects
are enhanced at the higher conductance values, where no suppression
of $\sigma_{S}$ at $2G_{0}$ is visible in Fig.~\ref{fig:AgAuPt-stddevSG}(a,b).
Also in the experiment\cite{Ludoph:PRB1999} only shallow minima have
been observed at the positions $nG_{0}$ for $n\geq2$. 

For the ESLM of a multivalent metal shown in Fig.~\ref{fig:sSGsimul}(d),
the first transmission channel is fully open only at $G=G_{0}/a_{1}\approx1.43G_{0}$,
so that a possible suppression of the fluctuations of $S$ is shifted
to conductance values above $1G_{0}$. However, the contribution of
the second partially open channel in the ESLM, and possibly further
channels in the experiments {[}see also Fig.~\ref{fig:AvChanT}(c){]},
strongly washes out the expected minimum at $G_{0}/a_{1}$. Together
with the incomplete opening of the dominant conduction channel {[}see
Fig.~\ref{fig:AvChanT}(c){]} these effects explain the absence of any
clear minima of $\sigma_{S}$ for the atomistic simulations in Fig.~\ref{fig:AgAuPt-stddevSG}(c).

\end{document}